\documentstyle[preprint,aps,prc]{revtex}
\input{psfig}
\tightenlines

\def\beq{\begin{equation}}
\def\eeq{\end{equation}}
\def\bea{\begin{eqnarray}}
\def\eea{\end{eqnarray}}
\def\eqref#1{Eq.~(\ref{eq:#1})}
\def\eqlab#1{\label{eq:#1}}

\def\VYP#1#2#3{{\bf #1} (#2) #3}  
\def\NP#1#2#3{Nucl. Phys. {\bf #1} (#2) #3}
\def\PL#1#2#3{Phys.~Lett. {\bf #1} (#2) #3}
\def\PR#1#2#3{Phys.~Rev.~{\bf #1} (#2) #3}

\def\ZP#1#2#3{Z.\ Phys.\  \VYP{#1}{#2}{#3} }
\def\half{\mbox{\small{$\frac{1}{2}$}}}

\newcommand{\vslash}[1]{#1 \hspace{-0.5 em} /}

\begin{document}
\title
{Production of e$^+$ e$^-$ pairs in proton-deuteron
capture to $^3$He  }
\author
{A. Yu. Korchin \footnote{Permanent address: National Science
Center `Kharkov Institute of Physics and Technology', 310108 Kharkov,
Ukraine}, D. Van Neck and M. Waroquier}
\address
{Department of Subatomic and Radiation Physics, University of
Gent, B-9000 Gent, Belgium}
\author
{O. Scholten and A. E. L. Dieperink}
\address
{Kernfysisch Versneller Instituut, 9747 AA Groningen, The
Netherlands}
\maketitle
\begin{abstract}
The process $p+d\leftrightarrow {^{3}\mbox{He}}+\gamma^*$ at intermediate 
energies is described using a covariant and gauge-invariant model, and 
a realistic pd$^3$He vertex. Both photodisintegration of $^3$He
and proton-deuteron capture with production of $e^+ e^-$  pairs are studied,
and results for cross sections and response functions are presented. 
The effect of time-like form factors on the dilepton cross sections is 
investigated as well.
\end{abstract}
\vspace*{\fill}
{{\em PACS}: 21.45.+v, 25.40.Lw, 25.20.Lj}
\newpage

\section{Introduction}

The electromagnetic probe is a well-established and powerful tool 
to investigate the structure of hadronic systems. In exclusive 
processes one can distinguish three different regimes depending on the 
four-momentum of the photon. First, in the space-like region ($q^2 < 0$) 
in quasi-free kinematics the $(e,e'p)$ reaction directly probes the 
single-particle structure of the nuclear vertex. Second, for real photons, 
the $(\gamma,p)$ or $(p,\gamma)$ reactions are more sensitive to the 
details of the reaction mechanism and to meson exchange currents (MEC).
Third, the rather less well-known time-like region ($q^2 > 0$), which can 
be explored using dilepton production, addresses additional aspects as
compared to 
real photons: (i) the coupling of longitudinally polarized  photons,
(ii) the time-like form factor in the ''unphysical'' region 
($4 m^2_e < q^2 < 4 m^2_p$).
Basically there are two alternatives to explore the time-like region:
virtual Compton scattering $p(\gamma,\gamma^*)p$ and the
bremsstrahlung processes with virtual 
photons, such as $p+p\rightarrow p+p+\gamma^*$ and capture reactions. 

In this paper we study the capture reaction 
$p+d\rightarrow {^{3}\mbox{He}}+\gamma^*$ at intermediate energies (proton 
energies up to a few hundred MeV) for both real and virtual photons. 
The main motivation for the present work are experiments at TSL (Uppsala) 
which in the past have explored \cite{Hoi} only small photon invariant masses 
($q^2 < $(10 MeV)$^2\,$) and KVI (Groningen) experiments which cover larger 
photon invariant masses \cite{Kal98}.  

Since we feel it is important to satisfy gauge invariance, we start from   
the covariant impulse appoximation model of \cite{Fal93} and make ~it 
explicitly gauge invariant by the introduction of an additional internal 
amplitude. An important input in this approach is the pd$^3$He vertex 
function, for which recent calculations \cite{Sch86,Wir95,For96} using a 
realistic NN interaction (Argonne v14 and v18) are used.

\section{Description of the model}

At small photon energies the photon production amplitude is dominated by 
radiation from the external legs (first three diagrams of Fig.~1), and
this consideration led to the development of low-energy theorems (LET)s 
\cite{Low58,Sak62,Gov81}. In kinematical conditions where the
photon energy is not small we may still assume the dominance of the
above external amplitude, without applying an expansion in powers of the 
photon energy. 

The 4-momenta of the proton, deuteron, $^3$He nucleus, and (virtual)
photon are denoted by  $p_1$, $p_2$, $p_3$, and $q$  respectively.
The external invariant amplitude reads as 
\bea
M_{ext}=e\,\epsilon_\mu^* (\lambda_\gamma )\,
\bar{u} (\vec{p}_3 ,\lambda_h )
\,M^{\mu\alpha}_{ext}\, u(\vec{p}_1 ,\lambda_p )\,\xi_\alpha (\lambda_d )\,,
\eqlab{1}
\eea
with
\bea
M^{\mu\alpha}_{ext}&=& 
\Phi^{\alpha}(p_3 ,p_2 ,p_1 -q) S(p_1 -q , m_1) \Gamma^{\mu} (p_1 -q
,p_1)\nonumber\\&& +
\Phi^{\beta}(p_3 ,p_2 -q ,p_1) \Delta_{\beta\rho}(p_2 -q )
\Gamma^{\rho\alpha\mu} (p_2 -q ,p_2 )\nonumber\\&&+ 
\Gamma^{\mu} (p_3 ,p_3 +q) S(p_3 +q , m_3)
\Phi^{\alpha}(p_3 +q ,p_2 ,p_1 )  \,,
\eqlab{2}
\eea
where $\epsilon_\mu^* (\lambda_\gamma )$ and $\xi_\alpha (\lambda_d )$
are the polarization vectors of the photon and the deuteron 
respectively,
$\bar{u}(\vec{p}_3 ,\lambda_h )$  ( $u(\vec{p}_1 ,\lambda_p )$ )
is  the 
spinor for $^3$He (proton) , $e$ is the proton charge, $S(k,m)$ is
the free propagator of the fermion  with mass $m$, 
$\Delta_{\beta\rho}(k)$ is the deuteron propagator. Helicities of the
particles are denoted by $\lambda$'s. 
The structure of the pd$\rightarrow
^3$He  vertex function $\Phi^{\alpha}$ will be discussed later. 

We stress here that in this formulation the $s$-amplitude (third diagram 
in Fig.~1) takes into account the pole contribution (but not the regular 
contribution) of the initial state pd interaction, and hence takes care 
of the problem with the orthogonality between initial and final state,  
mentioned in \cite{Nec94}. 

The electromagnetic (em) vertex function for spin-$\half$ 
particles  is chosen in the form
\beq
\Gamma^{\mu} (p -q,\,p)=\Gamma^{\mu} (p,\,p+q)
 =Z \gamma^\mu -
i\frac{\sigma^{\mu\nu}q_\nu}{2 m }F_2 (q^2)
+Z\tilde{F}_1 (q^2) (q^\mu \vslash{q} -q^2 \gamma^\mu ) \,,
\eqlab{4}
\eeq
where $F_1 (q^2)$ and $F_2 (q^2)$ are respectively the Dirac and Pauli
em form factor (FF), $\tilde{F}_1 (q^2) \equiv [1-F_1 (q^2)]/q^2\,$, and 
$Z=1\,(2)$ for the proton 
($^3$He). This vertex obeys the Ward-Takahashi identity for the
half-off-shell case \cite{Gro87}.

The half-off-shell $\gamma d d$ vertex satisfying the
corresponding Ward-Takahashi identity \cite{Sak62} can be written as 
\bea
&&\Gamma^{\rho\alpha\mu}(p_2 -q, p_2)=
 -g^{\rho\alpha}(2p_2 -q)^\mu +(p_2 -q)^\rho g^{\mu\alpha}
+\tilde{F}_1 (q^2 ) g^{\rho\alpha}[q^2 (2p_2 -q)^\mu  -q \cdot (2p_2-q)q^\mu ]
\nonumber\\ &&+F_2 (q^2 )(q^\rho g^{\mu\alpha}-q^\alpha g^{\mu\rho})
+\frac{F_3 (q^2 )}{2m_{2}^2}[\,q^\rho q^\alpha (2p_2 -q)^\mu -
\frac{1}{2}q\cdot (2p_2 -q)(q^\alpha g^{\mu\rho}+q^\rho g^{\mu\alpha})\,]\,,
\eqlab{15}
\eea
where $F_i (q^2)$ are related to the
charge $G_C(q^2)$, magnetic $G_M(q^2)$ , and quadrupole $G_Q(q^2)$ 
em FFs of the deuteron (see, e.g., \cite{Arn80}).

Apart from the amplitude $M_{ext}$ corresponding to radiation from the 
external legs there are other more complicated processes, such as 
initial-state pd rescattering and MEC. 
This contribution (henceforth called the internal amplitude $M_{int}$) can 
be constrained by imposing the gauge invariance requirement for the total 
amplitude $M=M_{ext}+M_{int}$. Such a  procedure is conventionally applied in 
derivations of the LET for bremsstrahlung \cite{Low58}. We will make use of
consequences of gauge invariance in situations where the photon energy is 
not small.

One can show that 
the internal amplitude obeys the following condition
\bea
q_\mu M_{int}^{\mu\alpha}=
-q_\mu M_{ext}^{\mu\alpha}=
\Phi^{\alpha}(p_3 ,p_2 ,p_1 -q )+\Phi^{\alpha}(p_3 ,p_2 -q ,p_1 )-
2\Phi^{\alpha}(p_3 +q ,p_2 ,p_1 )\,.
\eqlab{19}
\eea

The pd$^3$He vertex function  for the case where at 
most one particle is off its mass shell has the following structure 
\cite{Fal93}
\bea
\Phi^\alpha (k_3 ,k_2 ,k_1 )=
\phi^{\alpha}_+ (k_3 ,k_2 ,k_1 ) +
\phi^{\alpha}_- (k_3 ,k_2 ,k_1 )\frac{\vslash{k}_1 -m_1}{2m_1 } +
\frac{\vslash{k}_3 -m_3}{2m_3 }\phi^{\alpha}_- (k_3 ,k_2 ,k_1 )\,,
\eqlab{20}
\eea
where the last two terms correspond to negative energy states and only 
contribute when the proton or the helion are off their
mass shells.
We will use a form for the $\phi^{\alpha}_{\pm}$ which allows a direct 
relation with the nonrelativistic wave function (WF)
\bea
\phi^{\alpha}_{\pm} (k_3 ,k_2 ,k_1 )=
[\,\gamma^\alpha G_{\pm}(Q^2)-Q^\alpha H_{\pm}(Q^2)\, ]\gamma_5 \,,
\eqlab{21p}
\eea
where 
$Q^\alpha =\frac{M_r}{m_1} k_1^\alpha-\frac{M_r}{m_2} k_2^\alpha \,$
is the relative pd 4-momentum and 
$M_r =m_1 m_2/(m_1+m_2)$ is the reduced mass of the pd system.
For the $^3$He, proton and deuteron diagram in Fig.~1 the relative
momenta take the values
$\,Q_3^\alpha =\frac{M_r}{m_1} p_1^\alpha-\frac{M_r}{m_2} p_2^\alpha
\,$, $\,Q_1^\alpha=Q_3^\alpha-\frac{M_r}{m_1}q^\alpha$ and 
$Q_2^\alpha=Q_3^\alpha+\frac{M_r}{m_2}q^\alpha$ respectively. 

From Eqs. (\ref{eq:19}-\ref{eq:21p}) it follows that a solution for the 
internal 
amplitude can be constructed as $M_{int}^{\mu\alpha}=M^{\mu\alpha}_{int}(1)
+M^{\mu\alpha}_{int} (2)$, where 
\bea
M^{\mu\alpha}_{int}(1)&=&\{\,[\gamma^\alpha G_{-}(Q_1 ^2 )-Q_{1}^\alpha
H_{-}(Q_1 ^2 )]\frac{\gamma^\mu}{2m_1}
-\frac{\gamma^\mu}{m_3}  [\gamma^\alpha G_{-}(Q_3 ^2 )-Q_{3}^\alpha
H_{-}(Q_3 ^2 )]\,\} \gamma_5 \nonumber\\
&& +g^{\mu\alpha}
\,[\,\frac{M_r}{m_1}H_+ (Q_1 ^2 )-\frac{M_r}{m_2}H_+ (Q_2 ^2 )\,]\gamma_5\,,
\eqlab{26}
\eea 
\bea
&&M^{\mu\alpha}_{int} (2) = \frac{M_r}{m_1}(q-2p_1 )^\mu
R_1^\alpha + \frac{M_r}{m_2}(q-2p_2 )^\mu  R_2^\alpha +
\frac{M_r }{m_1+m_2 }(q+2p_3 )^\mu (R_1^\alpha+R_2^\alpha)\,.
\eqlab{37}
\eea
We have used the notations ( for i=1,2 )
\bea
R_i^\alpha =[\gamma^\alpha G_+'(Q_i^2) -Q_3^\alpha H_+'(Q_i^2)]\gamma_5\,
\,,\quad\quad\quad
G'_+ (Q_i^2)=\frac{G_+(Q_i^2)-G_+(Q_3^2)}
{Q_i^2-Q_3^2 }\,
\eqlab{35}
\eea
and similar notations for $H'_+(Q_i^2)$. 
Note, that this amplitude remains finite in the special cases where
$Q_1^2\rightarrow Q_3^2$ or  $Q_2^2\rightarrow Q_3^2$. As a check
of our results we verified that $M=M_{ext}+M_{int}$ reproduces the
LET amplitude of \cite{Gov81} for real photons when $q\rightarrow 0$. 

We include ~in the calculation the dominant 
components \cite{Sch86,For96} of the $^3$He WF, i.e. a $pn$ pair in the 
deuteron state or in the $^1S_0$ (quasi) bound $d^*$ state, coupled to a 
proton.  We neglect contributions to the amplitude where the deuteron 
is excited into the $T=0$ continuum.    
The $pd\rightarrow pd^*$ capture mechanism via the spin-flip 
$^3S_1 + ^3D_1\rightarrow ^1S_0$ transition has been shown \cite{Fal93}
to be important and is therefore included explicitly in $M_{ext}$.
The corresponding amplitude (Fig.~1, last graph) can be written as
 \bea
M^{\mu\alpha}_{d*} = \Psi (p_3, p_2 -q,p_1 )
\Delta (p_2 -q)
\Gamma^{\mu\alpha}(p_2 -q,p_2 )\,,
\eqlab{48}
\eea
where $\Delta(k)=(k^2-m^{*2}_{2}+i0)^{-1}$ and the em vertex
has the form
\bea
\Gamma^{\mu\alpha}(p_2 -q,p_2 )=-\frac{i}{m_1}\mu_v
\varepsilon^{\mu\alpha\rho\nu}q_\rho (p_2 )_\nu F(q^2)\,.
\eqlab{49}
\eea
Here  $\mu_v=\mu_p -\mu_n$ is the isovector magnetic moment of the
nucleon, $m^*_{2}$ is the mass of the $d^*$, $F(q^2)$ is the transition
FF and $\Psi(p_3, p_2 -q,p_1 )$ is the pd*$^3$He vertex function.
This contribution is gauge invariant and does not
affect the above discussion of the gauge invariance.

The invariant functions $G_\pm (Q^2)\,, H_\pm (Q^2)$ can be
related to the S and D components of the overlap integral 
$<d\,|\,^3 He>$. For this purpose the formalism developed previously 
in \cite{Buc79} for the pnd vertex and also in \cite{Fal93} has been applied.
In the same way the vertex
$\Psi (p_3, p_2 -q,p_1 )$ has been  expressed through the overlap integral
$<d^* \,|\,^3He>$. 
Two models for the $^3$He WF have been used in calculations.
The first one is the parametrization in \cite{Fal93} of the 
calculations in Ref. \cite{Sch86}
with the Argonne v14 NN + Urbana VII 3N interaction.
The second model is a more recent calculation
\cite{For96} with the Argonne v18 NN + Urbana IX 3N interaction. 

In order to calculate the em FFs of the proton 
we used the extended vector meson dominance  model \cite{VMD}.
The deuteron FFs at negative $q^2$ are
taken from the calculation in \cite{Wir95}, and we used the parametrization of
the FFs of the $^3$He as given in \cite{McC77}.
The proton FFs are continuous functions when going from negative
to positive $q^2$, and this behaviour is incorporated in the VMD
models.  For the deuteron and the helion
we have made the assumption that the em FFs have a smooth extrapolation
from the space-like to the time-like region. In this paper we are 
interested in the interval of relatively small photon invariant masses, 
restricted by the proton incoming energy $T_p$ of about 300 MeV. Since the 
maximal photon invariant 
mass $m_\gamma^{max}=\sqrt{s}-m_3 \approx 2/3 \, T_p\,$, $\,\,q^2$ does 
not exceed 0.04 GeV$^2$ and the above
approximation should give a reasonable estimate of the effect of FFs on the
dilepton cross sections. Finally, the FF of  
the transition $d \rightarrow d^*$ at positive $q^2$ is chosen the
same as the deuteron FF $F_1(q^2)$.

At  the real photon point the FFs are normalized to 
\bea
&&F_1^p (0)=F_1^h (0)=G_C(0)=1\,, \quad\quad
F_2^p (0)=\mu_p -1\,,\quad \quad F_2^h (0)=\frac{m_3}{m_1}\mu_h -2\,,
\nonumber\\
&&G_M (0)=\frac{m_2}{m_1}\mu_d \,,\quad \quad \quad G_Q (0)=m_2^2 Q_d\,,
\eqlab{42}
\eea
where $Q_d =0.2859$ fm$^2$ is the quadrupole moment of the deuteron and
the magnetic moments of the proton, deuteron, and $^3$He
(in nuclear magnetons) are  $\mu_p =2.7928$,  $\mu_d =0.85774$, and
$\mu_h = -2.12755$ respectively.

\section{Cross section and response functions for dilepton production}

The c.m.\  cross section for the p+d $\rightarrow\,^3$He+$e^++e^-$ reaction 
can be decomposed (in complete analogy to the spacelike $(e,e'p)$ reaction) 
into the sum of products of kinematical factors and four response functions
(RFs),
\bea
\frac{d\,\sigma \,(e^+e^-)}{d\,\Omega_\gamma d\,m_\gamma d\,\Omega_e^*}&=&
\frac{\alpha^2 m_1 m_3 q_c \beta}{16\pi^3 m_\gamma p_c s}\,
[\, W_T\, ( 1-\frac{1}{2}\beta^2 \sin^2 \theta^*)
+W_L\, ( 1-\beta^2 \cos^2 \theta^*) \nonumber\\
&&+W_{TT} \,\frac{1}{2}\beta^2 \sin^2 \theta^* \cos 2\phi^*
+W_{LT}\,  \frac{1}{2\sqrt{2}}\beta^2 \sin 2\theta^* \cos \phi^* \,]\,.
\eqlab{52}
\eea
Here $m_{\gamma}=\sqrt{q^2}$ is the invariant mass of the virtual photon, 
$s=(m_1 +m_2 )^2+2m_2 T_p\,$, $T_p$ is the proton kinetic energy in the lab 
frame, $p_c$ and $q_c$ are the c.m. 3-momenta of the proton (deuteron) and
photon (helion) respectively, $\alpha=1/137.035$,
$\beta= (1-4m_e^2/m_\gamma^2)^{1/2}$ and $m_e$ is the electron mass.
For a description of the kinematics and details about  
the decomposition of the $e^+ e^-$ cross section
into the independent RFs $W_i$ we refer to 
\cite{Nec94}. 
The differential $d\,\Omega_e^*$ in \eqref{52}
is written in the photon rest frame (denoted by $^*$) \cite{Nec94}, but can 
easily be transformed back to the proton-deuteron c.m. frame 
( see \cite{Kor95} ).

The RFs in \eqref{52} contain information on the hadronic transition. 
They depend on three variables 
($W_i \equiv W_i (s,m_{\gamma},\theta_{\gamma}$)) and are defined as 
\bea
&&W_T
=\frac{1}{6}\sum_{polar.}( |J_x|^2+|J_y|^2 )\,,\;\;\;\;\;\;\;\;\quad\quad
W_L
=\frac{1}{6} \frac{m_\gamma^2}{q_0^2}\sum_{polar.}|J_z|^2\,,\nonumber\\
&&W_{TT}
=\frac{1}{6}\sum_{polar.}( |J_y|^2-|J_x|^2 )\,,\;\;\;\;\;\;\quad\quad
W_{LT}
=-\frac{1}{6}\frac{m_\gamma}{q_0} \sum_{polar.}
2\sqrt{2}\,\Re\, (J_z J_x^* )\,,
\eqlab{53}
\eea
where $q_0=(m_\gamma^2+\vec{q}_c^2)^{1/2}$ is the energy of the virtual 
photon, and the space components of
the em current $J^\mu \equiv \bar{u} (\vec{p}_3 ,\lambda_h )
\,M^{\mu\alpha}\, u(\vec{p}_1 ,\lambda_p )\,\xi_\alpha (\lambda_d )\,$
are evaluated in the system with the OZ axis along the photon momentum. 
In obtaining these expressions gauge invariance has been used
to eliminate the time component of the current.

Integration of \eqref{52}  over the lepton angles leads to the expression
\bea
\frac{d\,\sigma \,(e^+ e^- )}{d\,\Omega_\gamma d\, m_\gamma }=
\frac{\alpha^2 m_1 m_3 q_c \beta
(1-\frac{1}{3}\beta^2) } {4\pi^2 m_\gamma p_c s }\,[\,W_T
(s,m_\gamma ,\theta_\gamma )
+ W_L (s,m_\gamma ,\theta_\gamma )\,]\,,
\eqlab{54}
\eea
where the interference RFs have dropped out.

An interesting observable is
the ratio of the $e^+e^-$ cross section (integrated over the
allowed photon invariant masses) to the real photon cross section
calculated at the same incoming energy and scattering angle. This
quantity is called the conversion factor, and as a ratio it is believed to be
less sensitive to many aspects of the reaction mechanism. We can cast
this ratio in the form $R(s,\theta_\gamma)=
R_T (s,\theta_\gamma)+R_L (s,\theta_\gamma)$,
where the transverse or longitudinal conversion factor is given by
\bea
R_{T,L} (s,\theta_\gamma ) = \frac{\alpha }
{\pi q_c'\, W_{T} (s,0,\theta_\gamma )  }
\int_{2m_e}^{m_\gamma^{max}}
{\beta (1-\frac{1}{3}\beta^2 )q_c W_{T,L} (s, m_\gamma, \theta_\gamma )\,
\frac{d\,m_\gamma}{m_\gamma} }\,.
\eqlab{57}
\eea
Here  $q_c=\{[s-(m_3+m_\gamma)^2][s-(m_3-m_\gamma)^2] \}^{1/2} /2
\sqrt{s}\, $, $\,\,q'_c=q_c\,|_{m_\gamma =0}$ 
stands for the real photon c.m. momentum and $m_\gamma^{max}=\sqrt{s}-m_3$.

\section{Results of calculations and discussion}
The model is first tested for the real photon reaction. 
Fig.~2 (upper panel) shows cross sections for the reaction 
$\gamma ^3$He$\rightarrow$pd at $E_{\gamma_{LAB}}$ = 245 MeV, related to the 
capture process at $T_{p_{LAB}}$ = 358 MeV via time reversal.
Note the unsatisfactory discrepancy between the 
disintegration and capture reaction measurements, known
for a long time and recently pointed out again in Ref. \cite{Saskat}.
As seen from the figure the agreement with the
photodisintegration data \cite{Bonn,Saclay} is quite reasonable and
considerably better than with the data \cite{TRIUMF} for the
capture reaction. The cross section here is determined mainly by the D 
component of the WF and differences between models 'a` 
( WF from \cite{For96} )  and 'b` ( \cite{Fal93} ) show up 
mainly at $\theta < 90^\circ$.  

To study the importance of the different contributions to the amplitude we
plot in Fig.~2 (lower panel) the energy dependence of the 
$\gamma ^3$He$\rightarrow$pd cross section at fixed angle. The figure 
shows in particular the large contribution of the internal amplitude (marked
'int` in Fig.~2 ). Also note that the cross section is a result of  
interference between all contribution, though the effect of the 
$^1S_0$ is relatively small. All calculations were performed in the Coulomb 
gauge (of course only the amplitude including the internal contribution is 
independent of the photon gauge). The full 
calculation (solid line) deviates from the
data at $E_\gamma$ between 100 and 300 MeV, though the disagreement is
not too large. We do not show results for model 'b`. In general, it
gives higher ( up to 40\% ) cross sections at small energies, while above 
50 MeV the situation is reversed.

We now discuss the $e^+e^-$ production in pd capture
and present calculations  at the proton lab energy 190 MeV (corresponding
to the kinematics at KVI, Groningen). 
The invariant mass and angular dependences of the response functions are 
shown in Fig.~3. The choice of a forward angle in Fig.~3 (left panel) was made 
because in these conditions the longitudinal response is enhanced. 
Both $W_L$ and $W_{LT}$ are comparable in magnitude to $W_T$. Fig. 3 shows 
the sensitivity of the RFs to such
ingredients of the model as the time-like FFs, the ``negative energy``
components $G_- ,\, H_-$ in the pd$^3$He vertex, and the $^1S_0$
contribution. The contribution of the components $G_- ,\, H_-$ is small, 
which can be explained by a cancellation between $M_{int}$ and the  
terms in $M_{ext}$ proportional to $\gamma^\mu$. 
Backward angles  are less favourable for  studying the longitudinal
response, which is almost always heavily suppresed compared to $W_T$.
The exception is at large $m_\gamma$, close to the kinematical limit.
For that reason we do not present the $m_\gamma$ dependence 
for backward angles.

Fig. 3 (right panel)  demonstrates the angular dependence at 
fixed $m_\gamma$= 65 MeV (which is about $\frac{1}{2}m_\gamma^{max}$).
The transverse RF has a dependence similar to that for the real photons.
The longitudinal RF is quite large at forward angles and diminishes
at backward angles. The interference RFs $W_{TT}$ and $W_{LT}$ also
show up in the forward hemisphere. The interference RFs vanish at 
$\theta_\gamma =0^\circ$ and $180^\circ$ because of the rotational symmetry
around the photon momentum in this kinematics.

From the experimental point of view the cross section integrated over
the photon invariant masses and the conversion factor are  of
considerable interest. These observables are plotted on Fig. 4.

Results for $R(s,\theta_\gamma)$
(Fig. 4 , lower left panel) are almost independent of the model for the WF, 
despite differences between the cross sections for these WFs
(Fig. 4 , upper left panel). The effect of the em FFs also turns
out to be very small in this kinematics, of the order of 1\%.
In general the longitudinal response is more sensitive to the FFs
(see also Fig.2); however, after integration over $m_\gamma$ the longitudinal 
cross section becomes extremely small
compared to the transverse one (Fig. 4, upper left panel) except at
very forward angles. This makes the study of the longitudinal response
in this integrated observable experimentally difficult.
On Fig. 4 we also show a model independent estimate for the conversion 
factor, which follows  from 
\eqref{57} by neglecting $W_L (s,m_\gamma, \theta_\gamma )$   
and the invariant mass dependence of the transverse response, i.e., assuming   
that $W_T(s,m_\gamma,\theta_\gamma) \approx W_T(s,0, \theta_\gamma)$.
On average, the deviation between the  $R(s,\theta_\gamma)$ in our model 
and the above estimate are of the order of 5\%.

At higher energy $T_p$ = 350 MeV ( $m_\gamma^{max}=229.8$ MeV ) 
one expects a larger influence of the FFs. The 
differences between the $^3$He models 'a` and 'b` in
the conversion factor are negligible and only calculations
with the model 'a` are presented. As it is seen, the FFs modify primarily
the longitudinal cross section, because the transverse one gets its main
contribution from the low $m_\gamma$ region, where $q^2$ dependence of
the FFs can be neglected. Compared to $T_p=190$ MeV, the weight
of the longitudinal cross section is now enhanced and as a result 
the conversion
factor is more sensitive to the FFs, though the effect is still not
larger than 5-7\%.

Note, that the time-like FFS of the deuteron and
$^3$He, when extrapolated from $q^2<0$ region,  rapidly 
increase with increasing $q^2$. 
At $\sqrt{q^2}=230$ MeV, for example, they are increased by a factor
two compared to 
their values at $q^2$=0. This may not be quite realistic, since 
the extrapolation does not take into account the off-shell effects in the    
em vertices and propagators. The general problem of the (half off-shell) 
time-like form factors for a weakly bound composite system like the deuteron 
or $^3$He is interesting, and will be addressed in future work.

In conclusion, a covariant and gauge invariant approach 
has been developed for the  
p+d $\leftrightarrow$ $^3$He+$\gamma^*$ reactions. The agreement with  
data for the $^3$He photodisintegration is reasonable, indicating that 
the approach seems to account for the basic mechanisms of
this process over a wide range of energies.
An important element of the approach is the internal amplitude needed
to ensure gauge invariance. The contribution of this part of the amplitude is
sizeable, which goes in line with the observations made in Ref. \cite{Nag93}.

A mechanism which is missing in this approach is the
initial- (or final-) state pd rescattering, although part of it is effectively
taken into account by the $^3$He diagram and the internal
contribution. The effects of explicitly 
including the pd interaction will be studied in a forthcoming 
publication \cite{fut}. 
Predictions have been made for the dilepton production experiments 
under way in Uppsala (TSL) and Groningen (KVI).
In general, the  longitudinal $W_L$ and the
interference $W_{LT}$ response functions  are more sensitive to the time-like
form factors. However, this effect can only be seen at forward angles
where $W_L$ and $W_{LT}$ are comparable in magnitude to $W_T$,
or at large angles and high photon invariant masses close to the
kinematical limit. Finally, we have calculated the conversion factor, which 
proves to be an almost model-independent observable.
\acknowledgements
We thank Robert Wiringa for calculating the $^3$He-d and $^3$He-d$^*$ overlap 
integrals. We would also like to thank Justus Koch, Ulla Tengblad, 
Jan Ryckebusch and Rob Timmermans for useful discussions, 
and Betsy Beise for sending data file with the deuteron form factors.
This work is supported by the Fund for
Scientific Research-Flanders (FWO-Vlaanderen).

\begin{figure}[1]
\vspace*{1cm}
\psfig{figure=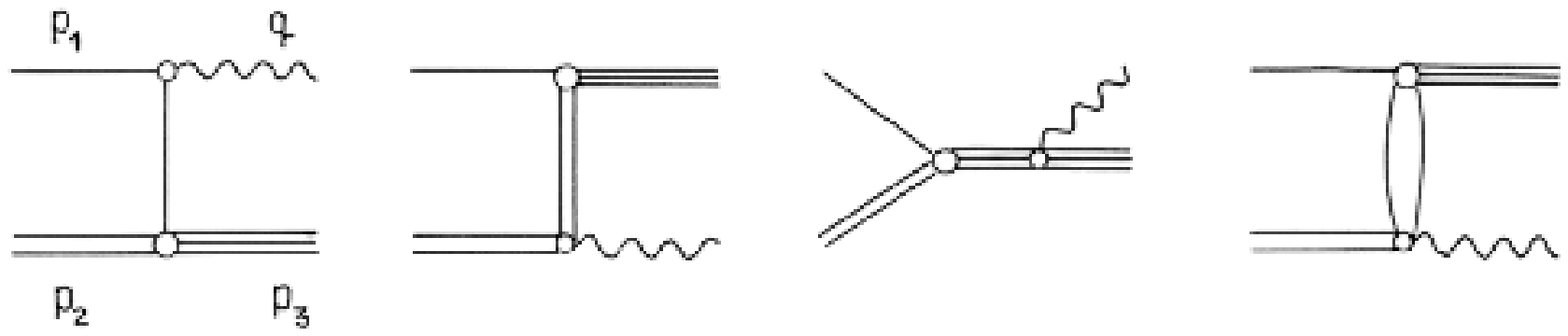,height=1.7in}
\caption[fig1]{Diagrams corresponding to the external amplitude.}
\label{fig:1}
\end{figure}

\begin{figure}[2]
\vspace*{1cm}
\psfig{figure=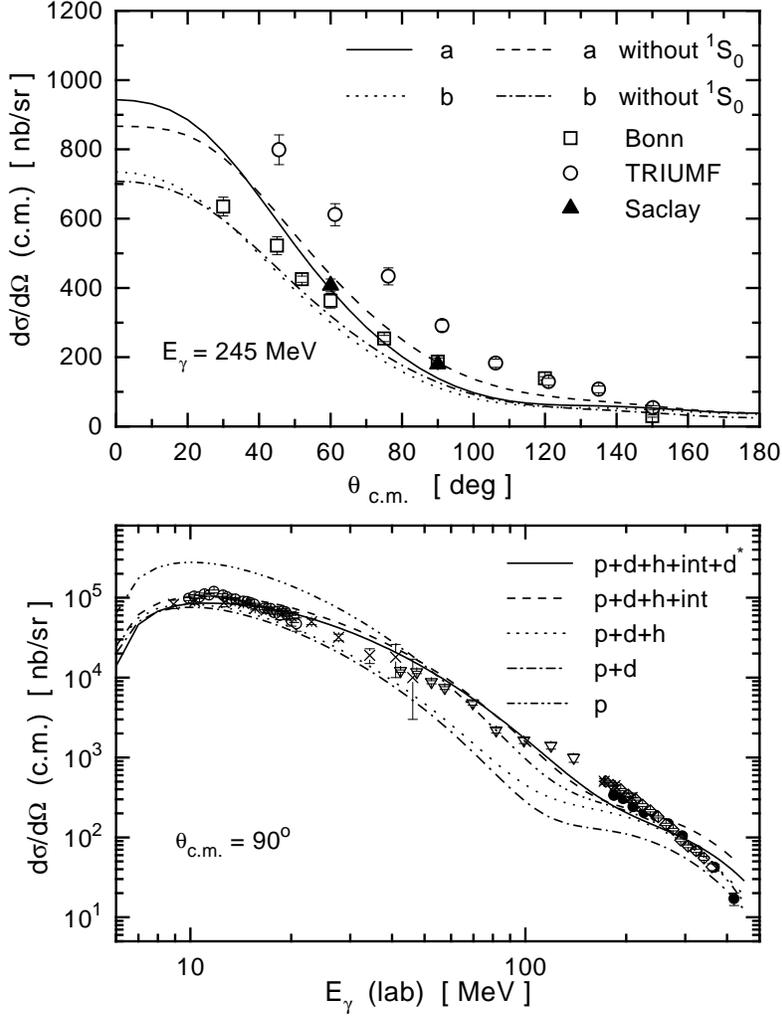,height=6.5in}
\caption[fig.2]{Cross section for the $\gamma ^3$He $\rightarrow$ pd reaction.
Upper panel: angular distribution, lower panel: energy dependence.
Curves on the upper panel are: solid and dashed ones for the model 'a`
        ( WF from~\protect\cite{For96} ), dotted and
         dash-dotted ones for the model 'b` ( WF
from~\protect\cite{Fal93} ).
Dashed and dash-dotted lines are calculations without $^1 S_0$
amplitude \eqref{48}, solid and dotted lines include all
contributions. The photodisintegration data are
taken from~\protect\cite{Bonn} and~\protect\cite{Saclay}. The TRIUMF
points are calculated from the capture data~\protect\cite{TRIUMF}
at $T_p=350$ MeV. Calculations shown on the lower panel are performed with 
model 'a`. The different curves correspond to calculations including 
different contributions to the reaction amplitude.
Data are from : $\diamond$~\protect\cite{Saclay}, 
 $\bigtriangledown$~\protect\cite{Illinois}, 
 $\bullet$~\protect\cite{Bonn}, $\circ$~\protect\cite{Cha74},
 $\times$~\protect\cite{Ste65}, 
 and $\ast$~\protect\cite{Saskat}.}
\label{fig:2}
\end{figure}

\begin{figure}[3]
\vspace*{1cm}
\psfig{figure=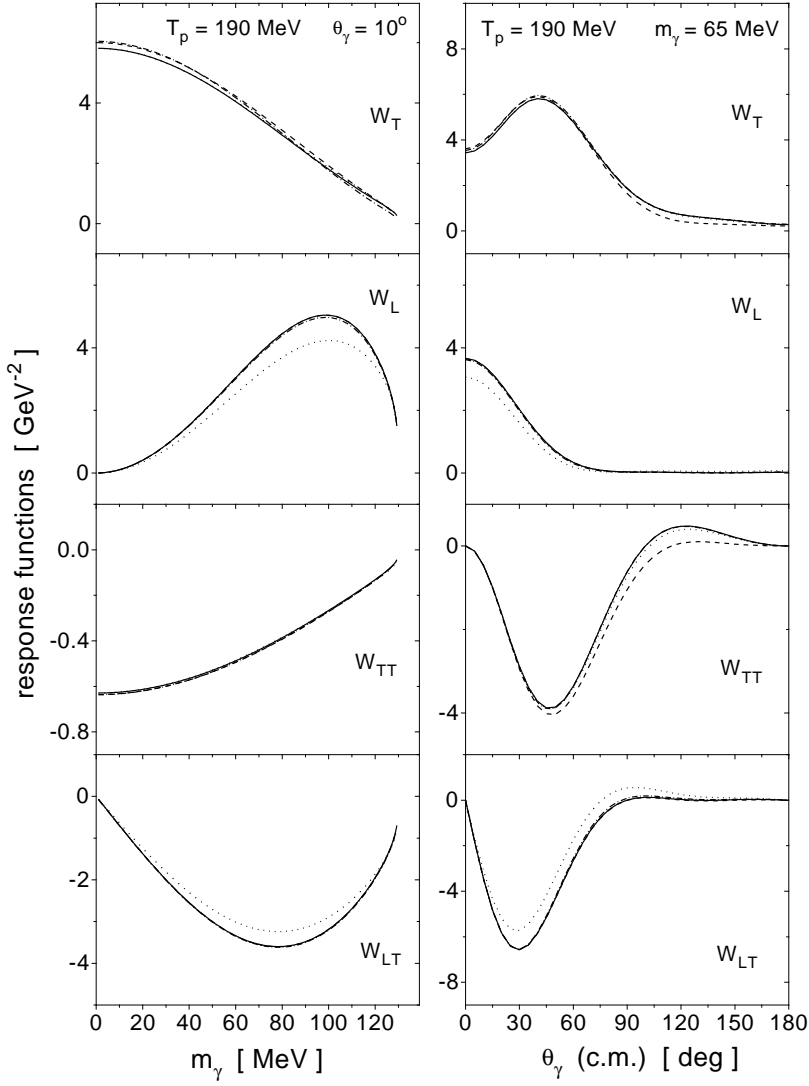,height=7in}
\caption[fig. 3 ]{Response functions for the $e^+e^-$ production in pd
capture for proton LAB energy 190 MeV. Left panel: photon
invariant mass distribution at $\theta_{\gamma}$= 10$^{\circ}$, right 
panel: angular distribution at
fixed $m_\gamma$=65 MeV.
Calculations are performed in the model 'a`.
 Solid lines are the full calculations, dashed ones are obtained
without the $^1S_0$ contribution. 
 Dotted lines are the results when the em FFs
 are switched off, dash-dotted lines are results  without the
$G_-,\,H_-$ components in the pd$^3$He vertex.}
\label{fig:3}
\end{figure}

\begin{figure}[4]
\vspace*{1cm}
\psfig{figure=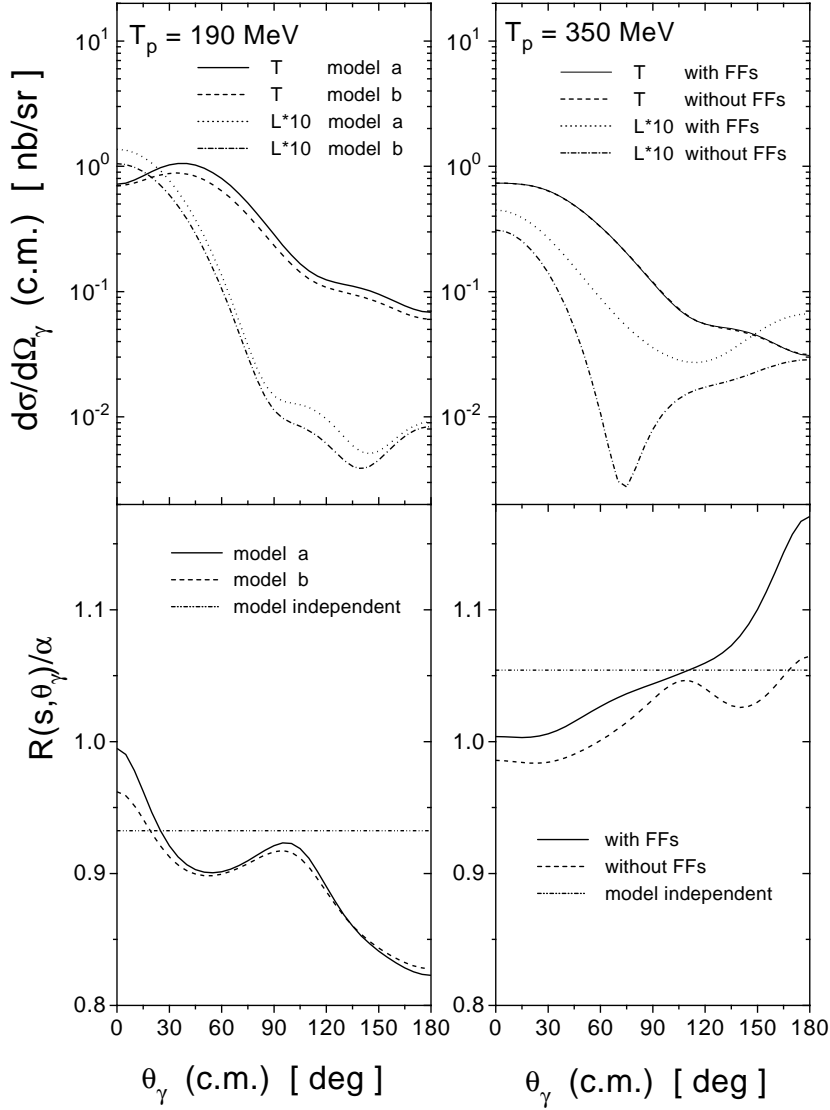,height=7in}
\caption[fig.4]{Angular dependence of the cross section integrated over 
$m_\gamma$ (upper panel) and the conversion factor devided by the
fine-structure constant (lower panel).
On the left panels the transverse (T) and longitudinal (L)
cross sections are calculated with different models for $^3$He WF.
The right panels show the effect of the form factors. The model independent
estimate for the conversion factor (see text) is shown by the  
dash-double-dotted lines.}
\end{figure}

\end{document}